 \documentclass[final,1p,times,]{elsarticle}

\journal{Physica A}
\usepackage[all]{nowidow}
\usepackage{csquotes}

\biboptions{sort&compress}

\newcommand{\EQ}[3]{
  \begin{equation}
    \label{#1}
    #2
    \;#3
  \end{equation}
}

\usepackage{mathtools}

\usepackage[detect-all]{siunitx}
\usepackage{bm}
\usepackage{amsmath}
\usepackage{bm}
\usepackage{xspace}

\begin{document}

\begin{frontmatter}



\title{Nonlinear diffusion of gluons}


\author{Georg Wolschin\corref{cor}}
\ead{g.wolschin@thphys.uni-heidelberg.de}
\address{Institut f{\"ur} Theoretische Physik der Universit{\"a}t Heidelberg, Philosophenweg 16, D-69120 Heidelberg, Germany, European Union}


\begin{abstract}
It is proposed to consider the fast thermalization of gluons in relativistic heavy-ion collisions as a diffusion process in momentum space. Closed-form analytical solutions of a nonlinear boson diffusion equation (NBDE) with constant drift and diffusion coefficients $\varv, D$ and boundary conditions at the singularity are derived.
The time evolution towards local central temperatures $T\lesssim 600$\,MeV through inelastic gluon scatterings 
in heavy-ion collisions is calculated for under- and overoccupied systems in the full momentum range. 
The results are consistent with QCD-based numerical calculations for gluon thermalization via inelastic gluon collisions.
\end{abstract}
 

\end{frontmatter}

\section{Introduction}
\label{intro}
Local equilibration in relativistic heavy-ion collisions has been investigated in great detail in the last 20 years.
Already in Ref.\,\cite{ba01}, thermalization has been considered to be the single most important question in the physics of heavy-ion collisions: It was not clear whether the system has enough time to equilibrate before falling apart. Meanwhile, there is ample evidence for fast thermalization from numerous investigations on the local equilibration of quarks and gluons, in particular, from effective kinetic theories that rely on the quantum Boltzmann equation such as Refs.\,\cite{km11,jpb12,jpb13,kur14,blmt17,fuku17}, and many others that are referenced therein. 
The thermalization time was found to be of the order of $\tau_\mathrm{eq}\lesssim 1$\,fm/$c$ \cite{kur14,fmr18}. The bottom-up scenario of \cite{ba01} was confirmed by solving the full effective kinetic theory of quantum chromodynamics (QCD) numerically \cite{kz15}.

These approaches rely on a weakly coupled description, in contrast to a strongly coupled paradigm built on the anti-de Sitter conformal field theory (AdS/CFT) correspondence \cite{son07} discovered in the investigation 
of D-branes in string theory \cite{mal98}, where partons are not the relevant degrees of freedom any more. In this work, it is assumed that quarks and gluons are well defined excitations in QCD at temperatures close to the critical value, and that kinetic theory is a valid tool to investigate partonic thermalization. 

The aim is to account for the time evolution during the fast thermalization of partons through a nonlinear diffusion equation for the occupation-number distributions in the full momentum range. The equation can be solved analytically for schematic initial conditions, whereas the quantum Boltzmann equation or other approximate kinetic equations \cite{jpb12,blmt17} must be solved numerically. It thus provides a transparent and analytically tractable model for aspects of the
kinetic evolution of a dense parton system during the early moments of a relativistic heavy-ion collision
 that were previously studied numerically in more involved approaches that were partly using QCD-related schemes such as the small-angle approximation for gluon scattering \cite{jpb13}.

For gluons, the model is based on a nonlinear boson diffusion equation (NBDE) that was proposed in Ref.\,\cite{gw18}, but now boundary conditions at the singularity $p=\mu$ with the chemical potential $\mu$ are considered, such that the approach offers closed-form solutions from the singularity at small momenta to the UV, albeit without uncovering the details of the underlying QCD processes that are hidden in the transport coefficients. In contrast, no singularity occurs for quarks where thermalization has already been studied within the corresponding nonlinear model through analytical solutions \cite{bgw19}. 

In accord with the available QCD-based numerical work such as \cite{jpb12,blmt17}, inelastic collisions are likely the main driver of thermalization. Consequently, the NBDE-model is solved in this work in closed form for inelastic collisions with $\mu=0$ that correspond to merging or splitting of gluons. Splitting moves particles towards lower energy, whereas merging moves them towards higher energies. The most important inelastic collisions are those where an extra gluon is produced or absorbed. For vanishing chemical potential $\mu=0$, the NBDE-solutions have a time-varying particle content until the stationary Bose--Einstein limit is reached at $t\rightarrow\infty$. Solutions for number-conserving elastic collisions are also possible, but require a time-dependent chemical potential $\mu(t)\le 0$ as investigated in the low-energy case of cold atoms \cite{gw20,sgw21}, where a Bose--Einstein condensate (BEC) is actually formed, and the calculated time-dependent condensate fraction is found to agree with data. Gluon condensate formation through elastic collisions appeared possible in overoccupied systems \cite{jpb12, jpb13}, but is found to be suppressed \cite{blmt17} because equilibration via inelastic scatterings occurs faster.

As in more involved numerical models of thermalization such as Refs.\,\cite{jpb12,blmt17}, a static plasma is considered that is spatially uniform and isotropic in momentum space. Obviously, the fireball in a relativistic heavy-ion collision does not fulfil these criteria in the course of its time evolution. However, gluon thermalization of the soft modes occurs on a very short timescale of $\tau_\mathrm{eq}\lesssim0.1$\,fm/$c$ compared to a collision duration of about $8-10$\,fm/$c$ in a central Pb--Pb collision at energies reached at the Large Hadron Collider (LHC). Hence,
the anisotropic -- preferentially Bjorken-like longitudinal, but also transversal -- expansion of the plasma that can be accounted for by hydrodynamics essentially starts only after the initial local equilibration and should therefore have no significant effect on the local thermalization. The isotropy in momentum space can be replaced by the assumption of isotropy in transverse momentum, which should be rather well fulfilled in central collisions, thus w.l.o.g. replacing $p\rightarrow p_\perp$ in the equations. 

The nonlinear diffusion model \cite{gw18} for quarks and gluons is briefly reviewed in the next section, but now including for gluons the most significant effect of the singularity on the analytical solutions through the boundary conditions. The fast thermalization of the soft modes via inelastic collisions is shown to occur as a consequence of the boundary conditions.
In Section\,3, the time-dependent solutions of the nonlinear boson diffusion equation in the full momentum range are calculated for schematic initial conditions via the logarithmic derivative of the time-dependent partition function, and the structure of Green's function with boundary conditions at the singularity is discussed. The results for the thermalization of the occupation-number distribution function through inelastic collisions are considered in Section\,4 for both under- and overpopulated systems. The conclusions are drawn in Section\,5.
\section{The model}
\label{model}
The simplest consideration of the time evolution during thermalization in the full momentum space is provided by the linear relaxation-time approximation with a momentum-independent relaxation time $\tau_\mathrm{rel}$. It leads to a fast equilibration not only in the low-momentum region, but simultaneously up to infinite momenta in the UV tail, which is also approached exponentially according to $\exp\,(-t/\tau_\mathrm{rel})$. Here, the $p=0$ mode is not reached instantaneously, but with the same finite time constant as the other modes.
In the relaxation ansatz, however, no account is made of the physical processes that cause thermalization.

For a more realistic description, one has to consider the interactions that mediate equilibration, and the inherent nonlinearity of the system.
A corresponding nonlinear partial differential equation
 for the single-particle occupation probability distributions $n\equiv n\,(\epsilon,t)$ had been derived
 from the quantum Boltzmann collision term with elastic collision kernel in Ref.\,\cite{gw18} as
 \begin{equation}
	\frac{\partial n}{\partial t}=-\frac{\partial}{\partial{\epsilon}}\left[\varv\, n\,(1\pm n)+n\frac{\partial D}{\partial \epsilon}\right]+\frac{\partial^2}{\partial{\epsilon}^2}\bigl [D\,n\,\bigr]
\label{nbde}
\end{equation}
where the $+$ sign represents bosons, and the $-$ sign fermions. Here, the new derivative-term of the diffusion coefficient is required 
such that the stationary solutions $n_\infty(\epsilon)$ become Bose--Einstein or Fermi--Dirac equilibrium distributions, respectively, as will be shown below. 

The model makes use of the ergodic approximation, where the occupation-number distributions 
$n\,(\epsilon,t)$ depend on energy $\epsilon$ and time. 
In the relativistic case, the energy dispersion relation is\footnote{~$\hbar=c=1$; time units are written as fm/$c$, momenta as GeV/$c$} $\epsilon=\sqrt{|\textbf{p}|^2+m^2}$ for massive particles, and
$\epsilon=|\textbf{p}|=p$ for massless particles such as gluons, whereas in the nonrelativistic case,  $\epsilon=|\textbf{p}|^2/(2m)$.             

When considering numerical solutions of the bosonic Boltzmann equation with elastic collision kernel, the thermal distribution with vanishing chemical potential is approached for $t\rightarrow \infty$ in the overpopulated case provided the boundary conditions at $\epsilon=0$ are imposed \cite{jpbwy14}; fermions attain the Fermi--Dirac distribution for $t\rightarrow \infty$. Correspondingly, the nonlinear diffusion model should also have the thermal distributions as a stationary limits.

To derive the stationary solution $n_\infty(\epsilon)$ of Eq.\,(\ref{nbde}) with
variable drift functions $\varv\,(\epsilon,t)$ and diffusion functions $D\,(\epsilon,t)$ that contain the many-body physics, rewrite Eq.\,(\ref{nbde}) and set the time derivative
to zero \cite{sgw21}
\begin{equation}
0=- \frac{\partial}{\partial \epsilon}{\left[\varv\,n_\infty \,(1\pm n_\infty)- D\,\frac{\partial n_\infty}{\partial \epsilon} \right]}\,,
\end{equation}
such that
\EQ{}
{\varv\,n_\infty(1\pm n_\infty) - D\,\frac{\partial n_\infty}{\partial \epsilon} = c_1}{.}
Dividing by $n_\infty(1\pm n_\infty)D$ and integrating over $\epsilon$ yields
\EQ{}
{
\int \frac{\text{d} n_\infty/\,\text{d} p}{n_\infty(1\pm n_\infty)}\,\text{d}\epsilon = 
\int \left( \frac{\varv}{D} - \frac{c_1}{n_\infty(1\pm n_\infty)D} \right)\text{d} \epsilon 
}{.}
Integrating the l.h.s. results in
$\left[\ln(n_\infty) - \ln(1\pm n_\infty)\right] = \ln\left (1-\frac{1}{1\pm n_\infty}\right)$ plus an integration constant $c_2$.
Solving for $n_\infty$ one obtains
\EQ{}
{n_\infty = \left[
\exp\left( 
\int  \left(- \frac{\varv}{D} + \frac{c_1}{n_\infty(1\pm n_\infty)D}\right)\text{d} \epsilon +c_2 \right) \mp 1
\right]^{-1}\,.
}{}
In order to reduce to the proper equilibrium distribution, $c_1 = 0$ is required, and the ratio $\varv(\epsilon,t)/D(\epsilon,t)$ must have no energy dependence for ${t\rightarrow\infty}$ so that it can 
be pulled out of the integral.
It follows that $\lim \limits_{t \to \infty}[-\varv\,(\epsilon,t)/D\,(\epsilon,t)] \equiv 1/T$ and $c_2\equiv -\mu/T$ such that the stationary distribution equals the Bose--Einstein/ Fermi--Dirac equilibrium distribution, respectively ($k_\text{B}=1$)
\begin{equation}
n_\infty(\epsilon)=n_\text{eq}(\epsilon)=\frac{1}{e^{(\epsilon-\mu)/T}\mp 1}
 \label{Bose--Einstein}
\end{equation}
with the chemical potential $\mu\le 0$ in a Bose system, and $\mu=\epsilon_\mathrm{F}>0$ for fermions. Here, the chemical potential appears as a parameter. For fermions, it is equal to the Fermi momentum, whereas for bosons, its initial value $\mu_\text{i}<0$ in case of elastic collisions is determined from particle-number conservation, and increases in time until the equilibrium value is attained --- or condensate formation starts when $\mu=0$ is reached in overoccupied systems. It vanishes in case of inelastic collisions that do not conserve particle number such that the temperature alone determines the equilibrium distribution. 

Hence, the NBDE Eq.\,(\ref{nbde}) can be applied not only to elastic, but -- setting $\mu=0$ -- also to inelastic gluon scatterings,
albeit with differing transport coefficients. The full complexity of the many-body system is hidden in the drift and diffusion functions, whereas the structure of  the nonlinear diffusion equation opens up the possibility for closed-form analytical solutions under the simplifying assumption of constant coefficients.

The drift and diffusion functions $\varv\,(\epsilon,t)$ and  $D\,(\epsilon,t)$ for an elastic collision kernel in the quantum Boltzmann equation are defined as first and second moments 
of the transition probabilities  $W_{\gamma \rightarrow \alpha}$ from state $|\gamma \rangle$ to $|\alpha\rangle$ \cite{gw18}
\begin{align}
W_{\gamma \rightarrow \alpha} = \sum_{\beta,\delta} \overline{V^2_{\alpha\beta\gamma\delta}} \,\mathcal{G}_{\alpha\beta\gamma\delta}\, (1+n_\beta)\,n_\delta
\end{align}	
with the second moment of the interaction $\overline{V^2_{\alpha\beta\gamma\delta}}$. The energy-conserving function 
$\mathcal{G}_{\alpha\beta\gamma\delta}$ 
has a finite width, but for an infinite system becomes proportional to 
$\delta\,(p_\alpha+p_\beta-p_\delta-p_\gamma)$ as in the usual Boltzmann approach. Introducing the density of states  $g_\alpha = g\,(\epsilon_\alpha)\propto \epsilon_\alpha^2$ for relativistic systems,
the sum can be replaced by an integral with $W_{\gamma \rightarrow \alpha}=W_{\gamma \alpha}\,g_\alpha$, and $W_{\alpha \rightarrow \gamma}$ accordingly.

The transport coefficients in the nonlinear diffusion equation thus depend on energy, 
 time, and the second moment of the interaction. They are integrals over distribution functions, and incorporate the nonperturbative many-body physics as accounted for by quantum chromodynamics. The drift term $\varv\,(\epsilon,t)$ has a negative sign and is mainly responsible for dissipative effects that drive the distribution towards the infrared and cause boson enhancement, the diffusion term $D\,(\epsilon,t)$ accounts for the diffusion of particles in momentum 
 space. In case of elastic collisions, this is reminiscent of classical particle diffusion with the Einstein relation connecting drift and diffusion, whereas for interactions that do not conserve the particle number the fluctuation-dissipation relation becomes more complicated. 
  
The nonlinear diffusion equation for the occupation-number distribution $n\,(\epsilon,t)$ becomes particularly simple when we assume energy-independent transport coefficients
\begin{equation}
	\frac{\partial n}{\partial t}=-\varv\,\frac{\partial}{\partial{\epsilon}}\Bigl[n\,(1\pm n)\Bigr]+D\,\frac{\partial^2n}{\partial{\epsilon}^2}.
	\label{bose}
\end{equation}
As in the general case before, it is straightforward to show that the stationary solutions are given by the thermal distributions for bosons and fermions.
The equation with constant transport coefficients still preserves the essential features of Bose--Einstein/Fermi--Dirac statistics which are contained in the quantum Boltzmann equation. This refers especially to the Bose enhancement in the low-momentum region that increases rapidly with time. Indeed, for ultracold bosonic atoms it has been shown \cite{sgw21} that the simplified equation with constant transport coefficients already accounts for evaporative cooling, and time-dependent condensate formation through elastic collisions with a time-dependent chemical potential $\mu(t)\le 0$ that enters the problem via the requirement of atom-number  conservation at every timestep with the appropriate density of states. In agreement with the available data for bosonic atoms, time-dependent condensate formation starts when $\mu=0$ is reached, and the subsequent time evolution of the condensate fraction can be properly accounted for. 

The diffusion equation with constant coefficients can be solved in closed form using the nonlinear transformation outlined in Ref.\,\cite{gw18} for any given initial condition $n_\mathrm{i}\,(\epsilon)$. For bosons, however,
it is more difficult to solve analytically due to the singularity at the chemical potential $\mu\le 0$, and the need to consider the boundary conditions at the singularity. No corresponding singularity exists for fermions, such that the exact solution of the nonlinear problem can be obtained with the free Green's function \cite{gw18,bgw19}. 

In this work, I focus on the thermalization of massless gluons with $\epsilon=|\textbf{p}|=p$ as accounted for by the nonlinear boson diffusion equation (NBDE) with constant coefficients, emphasizing inelastic collisions which substantially violate particle-number conservation through either splitting or merging processes. While number-conserving elastic collisions in Bose systems could mediate the formation of a Bose--Einstein condensate also in case of gluons as proposed in Ref.\,\cite{jpb12}, the violation of gluon-number conservation in relativistic collisions essentially precludes condensate formation \cite{blmt17} -- which could otherwise occur in overoccupied systems \cite{jpb12} where the initial gluon content is larger than that of the final equilibrium distribution.

The NBDE solutions for fixed chemical potential $\mu=\text{const}$ have the correct Bose--Einstein equilibrium limit for $t\rightarrow \infty$.
 They do not conserve the particle number \cite{sgw21}
\begin{equation}
	N(t)=\int_0^\infty g\,(\epsilon)\,n\,(\epsilon,t)\,\text{d}\epsilon\ne \text{const}\,,
	\label{ntot}
\end{equation}
except for systems that have a critical initial occupation at the boundary from an under- to an overoccupied situation.
Consequently, the solutions can account for number-violating inelastic collisions with $\mu=0$ that lead to local statistical equilibrium in both, under- and overoccupied systems --- even though gluon splitting or fusion is not considered explicitly in the NBDE, but instead hidden in the transport coefficients.

Number-conserving elastic collisions have been widely discussed in the literature based on QCD-inspired numerical solutions of Boltzmann-type transport equations for gluons, see Blaizot et al. \cite{jpb12,jpb13}, where the possibility of gluon-condensate formation in overoccupied systems through elastic collisions was first introduced. In the present context of a nonlinear diffusion equation, elastic collisions with $\mu_\text{i}<0$ require a time-dependent chemical potential
as obtained from particle-number conservation. This allows to calculate the time-dependent condensate fraction --- as detailed in Ref.\,\cite{sgw21} for ultracold atoms, where condensation actually occurs.

For inelastic collisions and in the simplified case of constant transport coefficients, I consider the drift and diffusion coefficients for gluons as parameters. Their values are determined from the respective relations to the equilibrium temperature $T$, and the local equilibration time $\tau_\mathrm{eq}\equiv\tau_\infty(p=Q_\mathrm{s})=4D/(9\varv^2)$ derived in Ref.\,\cite{gw18} at the gluon saturation  momentum $p=Q_\mathrm{s}$  as
\begin{equation}
 D=4\,T^2/(9\tau_\mathrm{eq}), \hspace{0.6cm} \varv=-4\,T/(9\tau_\mathrm{eq})\,.
 \label{Dv}
 \end{equation}
The relation between the transport coefficients and the local equilibration time arises from an asymptotic expansion of the error functions in the analytical solutions \cite{gw18} at the UV boundary $p=Q_\mathrm{s}$, whereas $T=-D/\varv$ is a consequence of the equality of the stationary solution with the Bose--Einstein distribution.  The values of equilibrium temperature and equilibration time are estimated using experimental information from the LHC, with $T\simeq600$\, MeV \cite{hnw17} and 
$\tau_\mathrm{eq}\le 0.25$\,fm/$c$. Here, the upper limit $\tau_\mathrm{eq}=0.25$\,fm/$c$ from Ref.\,\cite{fmr18},  corresponds to a fixed average value in a relaxation-time approximation of their coupled kinetic equations. Such a method enforces a fast exponential approach of the Maxwell--Boltzmann tails and hence, favours long relaxation times. In the present nonlinear approach, the equilibration time must therefore be shorter, and I shall choose $\tau_\mathrm{eq}\simeq 0.14$\,fm/$c$, see Section 4. 

With the corresponding values for the drift and diffusion coefficients, the time evolution towards equilibrium is determined through the solution of the NBDE in the full momentum range. Of course, the transport coefficients would eventually differ for elastic and inelastic collisions.

The solution $n\,(\epsilon,t)$ of the nonlinear boson diffusion equation without boundary conditions \cite{gw18} ist first briefly reviewed.
It can be written as
\begin{equation}
	n\,(\epsilon,t) =T \frac{\partial}{\partial{\epsilon}}\ln{\mathcal{Z}(\epsilon,t)} -\frac{1}{2}= T\frac{1}{\mathcal{Z}} \frac{\partial\mathcal{Z}}{\partial{\epsilon}} -\frac{1}{2}
	\label{eq:Nformula}	
\end{equation}
with the time-dependent partition function ${\mathcal{Z}(\epsilon,t)}$ 
     \begin{equation}
    \mathcal{Z}(\epsilon,t)= \int_{-\infty}^{+\infty}{G}\,(\epsilon,x,t)\,F\,(x)\,\text{d}x
    \label{eq:partitionfunctionZ}
    \end{equation}
which is an integral over Green's function $G\,(\epsilon,x,t)$
of the linear diffusion equation    
  \begin{equation}
   \left[ \frac{\partial}{\partial t}- D \frac{\partial^2}{\partial \epsilon^2}\right]G\,(\epsilon,x,t)=\delta(\epsilon-x)\,\delta(t)
    \label{eq:diffusionequation}
\end{equation}
and an exponential function that contains a definite integral over the initial nonequilibrium gluon distribution $n_\text{i}$ \cite{gw18}
\begin{equation}
	F\,(x) = \exp\,\left[-\frac{1}{2D}\left( \varv x+2\varv\int_0^xn_\text{i}(y)\,\text{d}y \right)\right]\,.
	\label{fini}
\end{equation}
When taking the logarithmic derivative to obtain the single-particle distribution function $n\,(\epsilon,t)$ from Eq.\,(\ref{eq:Nformula}), 
the value of the definite integral in Eq.\,(\ref{fin}) over the initial distribution $n_\mathrm{i}$ at the lower boundary $y=0$ cancels out. We can therefore replace the definite integral without loss of generality by an indefinite integral with an integration constant set to zero, 
\begin{equation}
\int_0^x n_\text{i}(y)\,\text{d}y \rightarrow \int^xn_\mathrm{i}(y)\,\text{d}y\,.
	\label{fin}
\end{equation}
This replacement considerably facilitates the explicit analytical integrations.
For a solution of the diffusion problem without boundary conditions, Green's function \(G_\mathrm{free}(\epsilon , x , t)\) of Eq.\,(\ref{eq:diffusionequation}) is a single Gaussian
\begin{equation}
	G_\mathrm{free}(\epsilon,x,t)=\frac{1}{\sqrt{4\pi Dt}}\,\exp\left[- \frac{(\epsilon-x)^2}{4Dt}\right]\,.
	\label{eq:Greensnonfixed}
\end{equation}
In Ref.\,\cite{gw18}, the occupation-number distribution had been obtained analytically with the free Green's function and the nonlinear transformation, 
Eq.\,(\ref{eq:Nformula}). The energy-independent prefactor $1/\sqrt{4\pi Dt}$ drops out when taking the logarithmic derivative. Although the solution that is obtained with the free Green's function is not only mathematically, but also physically correct in the UV, it causes a depletion of occupation in the IR due to diffusion into the negative-energy region --- and consequently, it does not attain the Bose--Einstein limit at low energies.

As a remedy, I now consider boundary conditions for bosons at the singularity $\epsilon=\mu=0$ for inelastic collisions.
This requires a new Green's function that equals zero at \(\epsilon=0\) $\forall \,t$
\begin{equation}
	G_\text{b}\,(\epsilon,x,t) = G_\mathrm{free}(\epsilon,x,t) - G_\mathrm{free}(\epsilon,-x,t)\,.
	\label{Greens}
\end{equation}
With this  bounded  Green's function, the time-dependent partition function becomes zero at the singularity for all times,
$\mathcal{Z}_\text{b}(\epsilon=0,t)=0$. Correspondingly, the occupation-number distribution that is calculated from the nonlinear transformation Eq.\,(\ref{eq:Nformula}) becomes infinity as the energy approaches zero from above,  \(\lim_{\epsilon \downarrow 0} n\,(\epsilon,t) = \infty\) \,$\forall$ \(t\), and it attains the Bose--Einstein limit over the full energy range as $t\rightarrow \infty$, not only in the thermal tail as in case of the free solutions.
 Moreover, the energy range is now restricted to $\epsilon \ge 0$, and the time-dependent partition function that includes the boundary conditions becomes for inelastic collisions
      \begin{equation}
    \mathcal{Z}_\text{b}(\epsilon,t)= \int_0^{+\infty}{G}_\text{b}\,(\epsilon,x,t)\,F\,(x)\,\text{d}x\,.
    \label{eq:partitionfunctionZb}
    \end{equation}
For any given initial nonequilibrium distribution $n_\mathrm{i}$, we can now solve the NBDE including boundary conditions at the singularity with the solution given by Eq.\,(\ref{eq:Nformula}) and $\mathcal{Z}\rightarrow    \mathcal{Z}_\text{b}$.  In case of inelastic scatterings with $\mu=0$, the particle number is not conserved, thermalization occurs mostly due to gluon splittings in the underoccupied case, and gluon fusion in the overoccupied situation.


\section{Thermalization of massless gluons}
\begin{figure}[t!]
	\centering
	\includegraphics[scale=0.7]{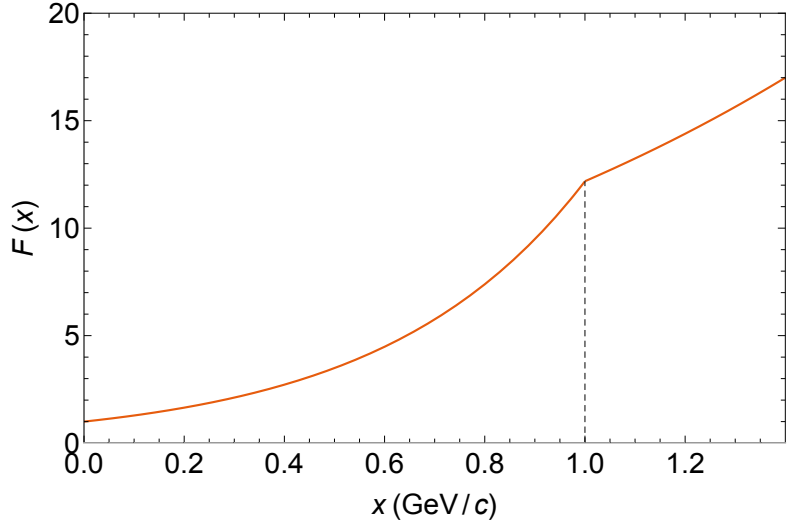}
	\caption{
		The function $F\,(x)$ of Eq.\,(\ref{fxmu1}). It contains the integral over an initial nonequilibrium gluon distribution $n_\mathrm{i}(x)$  with average initial occupation $n_\mathrm{i}^0=1$ according to Eq.\,(\ref{inix}). The transport coefficients $D, \varv$ are given in the text.
	}
	\label{fig1}
\end{figure}
\begin{figure}[t!]
	\centering
	\includegraphics[scale=0.7]{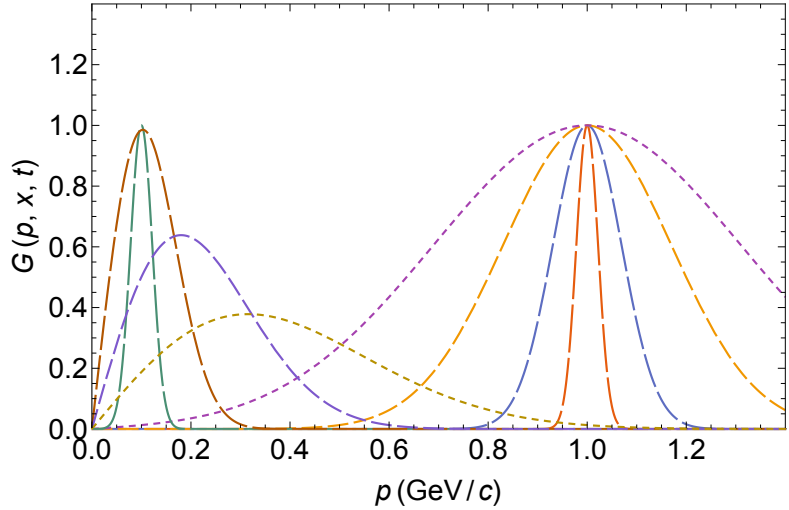}
	\caption{
		Green's function $ {G}_\text{b}(p,x,t)$ of Eq.\,(\ref{bgreens}) without the momentum-independent prefactor for $x=0.1$\,GeV/$c$ (left curves) and $x=1$\,GeV/$c$ (right curves) at $t=0.0002,  0.002, 0.012$ and $0.04$\,fm/$c$ (decreasing dash lengths) with $\mu=0$. As required by the boundary conditions at the singularity, $ {G}_\text{b}(p=\mu,x,t)=0~ \forall t$. The transport coefficients are as in Fig.\,\ref{fig1}.		
	}
	\label{fig2}
\end{figure}
For massless gluons, we have $\epsilon=|\textbf{p}|=p$. At the onset of a relativistic hadronic collision, an initial-momentum distribution $n_\mathrm{i}(|\mathbf{p}|)\equiv\,n_\mathrm{i}(p)$ has been used by Mueller \cite{mue00} based on the McLerran--Venugopalan model \cite{mlv94}.
It accounts, in particular, for the situation at the start of a relativistic heavy-ion collision \cite{jpb12}. This amounts to assuming that all gluons up to a limiting momentum $Q_\mathrm{s}$ are freed on a short time scale $\tau_0\sim Q_\mathrm{s}^{-1}$, whereas all gluons beyond $Q_\mathrm{s}$ are not freed.
Thus the initial gluon-mode occupation in a volume $V$ is taken to be a constant up to $Q_\mathrm{s}$.
Typical gluon saturation momenta for a longitudinal momentum fraction carried by the gluon $x\simeq 0.01$ turn out to be of the order $Q_\mathrm{s}\simeq 1$\,GeV \cite{mtw09}, which is chosen for the present model investigation.
Hence, the schematic initial conditions are taken as
\begin{equation}
	n_\mathrm{i}\,(p,t=0)=n_\text{i}^0\,\theta\,(1-p/Q_\mathrm{s})
	\label{inix}
\end{equation}
with an average initial occupation $n_\text{i}^0$. If the chemical potential vanishes as in case of inelastic collisions, the equilibrium temperature $T$ and the initial occupation were found to be related 
in Ref.\,\cite{jpb13} 
as $T=[15n_\text{i}^0/(4\pi^4)]^{1/4}Q_\mathrm{s}$, yielding $T=600$\,MeV
for $Q_\mathrm{s}=1$\,GeV$/c$ and $n_\text{i}^0\simeq 3.37$ assuming conserved energy density. This corresponds to an overpopulated system where the total particle number decreases during the time evolution, essentially through gluon merging via inelastic collisions \cite{jpb12,blmt17}. For a theta-function initial distribution, the boundary between under- and overpopulated systems is at $r_\mathrm{c}=n_\text{i}(0)/n_\text{eq}(T\ln 2)\simeq 0.154$ \cite{jpb13}. The NBDE-solutions with $\mu=0$ indeed fulfil particle-number conservation for this critical initial occupation.

In the model calculations presented in this work the focus is on inelastic collisions, because these will be decisive for the rapid thermalization.  We shall first consider an underpopulated system of arbitrary initial occupation $n_\text{i}^0\simeq 0.12$ -- below the corresponding critical value of  $r_\mathrm{c}=0.154$ -- that eventually approaches an equilibrium temperature $T=267$\,MeV, mainly through gluon splittings.
This is later compared with a calculation for $n_\text{i}^0=3.37$ and $T=600$\,MeV as an example for a physically realistic overpopulated case: Such a temperature is in the range of what is expected for the initial central temperature in heavy-ion collisions at energies reached at the LHC \cite{hnw17}, before hydrodynamic expansion and cooling of the initially produced fireball set in.


The time-dependent partition function $ \mathcal{Z}_\text{b}(\epsilon\equiv p,t)$ from Eq.\,(\ref{eq:partitionfunctionZb}) can be calculated as an integral over the bound Green's function $G_\text{b}(p,x,t)$ from
Eq.\,(\ref{Greens}) times the function $F\,(x)$ from Eq.\,(\ref{fini}) that contains the above initial gluon distribution.
The indefinite integral in Eq.\,(\ref{fini}) over the initial condition Eq.\,(\ref{inix}) can be carried out to obtain 
\begin{eqnarray}
	\label{fxmu1}
		F\,(x)=\exp\left[-\varv\,x/(2D)-(\varv\,n_\text{i}^0/D)(Q_\mathrm{s}-x)\,\theta\,(x-Q_\mathrm{s})+x\right]\,.
\end{eqnarray}
The function $F\,(x)$ is continuous, but not differentiable at $x=Q_\mathrm{s}$. This is essential to account for the equilibration in the UV region. $F\,(x)$ is plotted in Fig.\,\ref{fig1} for $n_\text{i}^0=1$ and the transport coefficients $D=1.2$\,GeV$^2/$fm and $\varv=-2$\,GeV$/$fm such that $T=-D/\varv=600$\,MeV.

Green's function of Eq.\,(\ref{Greens}) that includes the IR boundary condition for inelastic collisions becomes
\begin{equation}
	\label{bgreens}
	G_\text{b}\,(p,x,t)=\frac{1}{\sqrt{4\pi Dt}}\left[\exp\left(\frac{-(p -x)^2}{4Dt}\right)
-\exp\left(\frac{-(p+x)^2}{4Dt}\right)\right]\,.
\end{equation}
It is plotted in Fig.\,\ref{fig2} as function of momentum $p$, using the same transport coefficients as in Fig.\,\ref{fig1}.  Here, the momentum-independent prefactor that cancels out when taking the logarithmic derivative is omitted. 
Results are shown
at $x=0.1$\,GeV/$c$ and $x=1$\,GeV/$c$ for four values of time, $t=0.0002, 0.002, 0.012$ and $0.04$\,fm/$c$ (decreasing dash lengths). It vanishes at the singularity in momentum space, ${G}_\text{b}\,(p=0,x,t)=0~ \forall t$ as required by the boundary conditions, and thus, causes the system to thermalize instantaneously at $t=0$. For elastic collisions with 
$\mu_\mathrm{i}<0$, the thermalization time at $p=0$ would be larger 
because it takes time for the chemical potential to reach zero, and -- in the overoccupied case -- for the condensate fraction to attain its equilibrium value.

With $F\,(x)$ and $G_\text{b}\,(p,x,t)$, the time-dependent partition function $\mathcal{Z}_\text{b}(p,t)$ and its derivative $\partial\mathcal{Z}_\text{b}/\partial{p}$ can now be calculated, as well as the occupation-number distribution $n\,(p\equiv\epsilon,t)$ from Eq.\,(\ref{eq:Nformula}). The full calculation may be carried out analytically in order to obtain an exact solution of the nonlinear problem, as was done in Ref.\,\cite{gw18} for the free case. When including now the boundary conditions at the singularity, 
I compute the partition function and its derivative using the \texttt{NIntegrate} and \texttt{Derivative} routines of Mathematica instead.

\section{Discussion of the solutions for inelastic collisions}

\begin{figure}[t]
	\centering
	\includegraphics[scale=0.80]{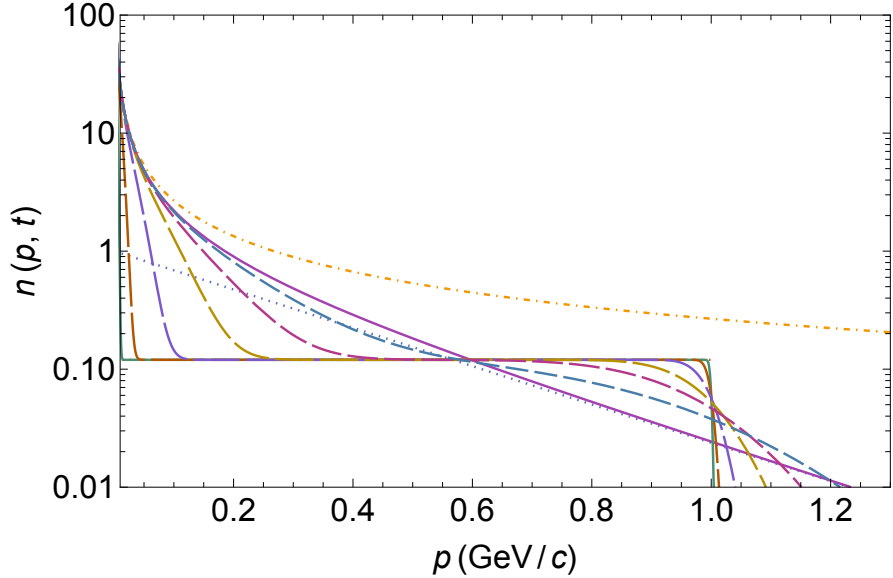}
	\caption{\label{fig3}
		Local thermalization of gluons in an underoccupied system with $n_\text{i}^0=0.12$ through inelastic collisions as represented by time-dependent solutions of the NBDE for $\mu=0$. Starting from schematic colour-glass initial conditions 
		Eq.\,(\ref{inix}) in the cold system at $t=0$, a Bose--Einstein equilibrium distribution with temperature $T=267$\,MeV (solid curve) is approached. The transport coefficients are $D = 0.23$\,GeV$^2$/fm, 
$\varv=-0.86$\,GeV/fm. Time-dependent single-particle occupation-number distribution functions are shown at $t =2\times10^{-5}, 2\times10^{-4}, 2\times10^{-3}, 0.01, 0.04$ and $0.12$\,fm/$c$ (decreasing dash lengths). 
Equilibrium is reached instantaneously for $p=0$ and is approached from below in the IR. The dot-dashed curve is the Rayleigh--Jeans, the dotted line the Maxwell--Boltzmann distribution.	
	}
\end{figure}
\begin{figure}[t]
	\centering
	\includegraphics[scale=0.80]{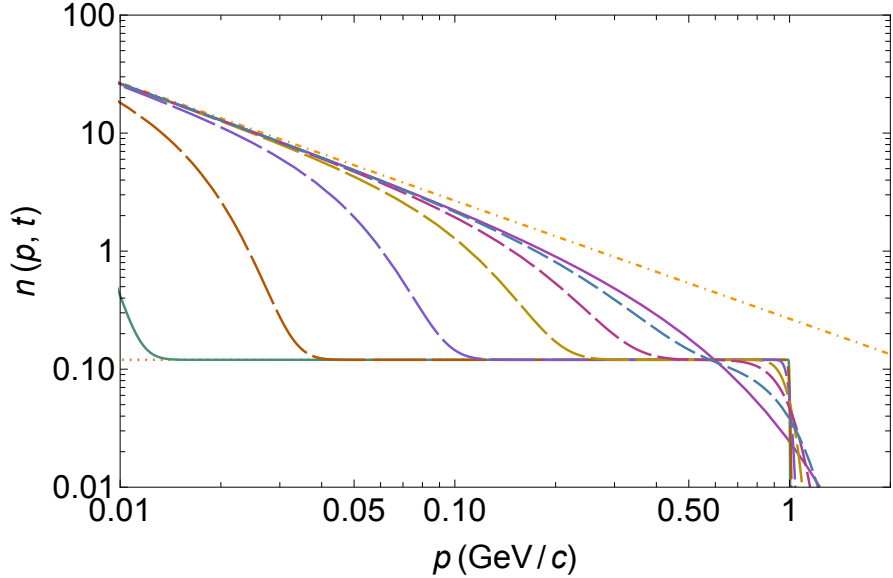}
	\caption{\label{fig4}
		Same calculation as in Fig.\,\ref{fig3} for inelastic collisions, but in a log-log-plot to emphasize the equilibration of the soft modes (underoccupied case).
		Starting from initial conditions Eq.\,(\ref{inix}) with $n_\text{i}^0=0.12$, a Bose--Einstein equilibrium distribution with temperature $T=267$\,MeV (solid upper curve) is approached. Time-dependent single-particle occupation-number distributions are shown at $t =2\times10^{-5}, 2\times10^{-4}, 2\times10^{-3}, 0.01, 0.04$ and $0.12$\,fm/$c$  (decreasing dash lengths). The dot-dashed line is the Rayleigh--Jeans spectrum.		
	}
\end{figure}
\begin{figure}[t]
	\centering
	\includegraphics[scale=0.802]{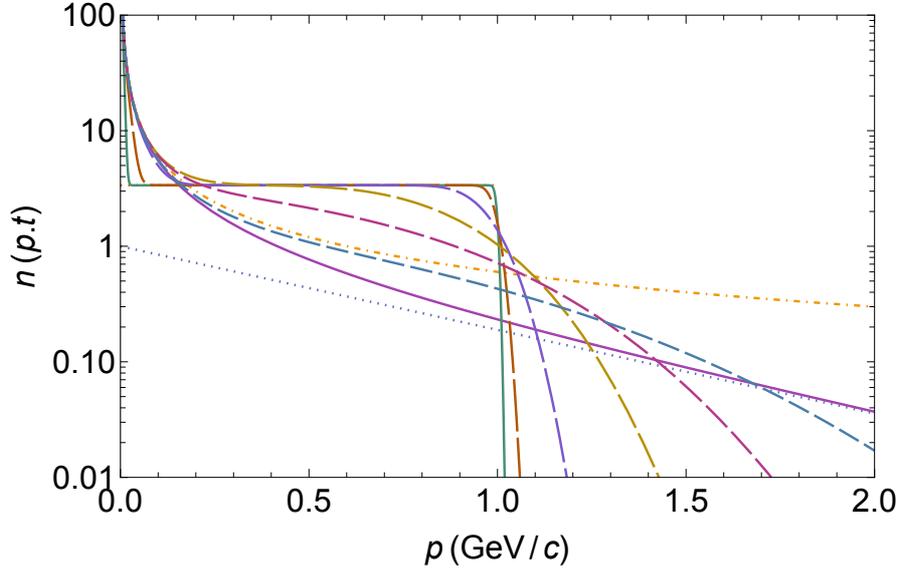}
	\caption{\label{fig5}
		Thermalization of gluons through inelastic collisions in an overoccupied case with the same time sequence as in Fig.\,\ref{fig3}. The initial occupation is enhanced to $n_\text{i}^0=3.37$ corresponding to a final equilibrium temperature $T=600$\,MeV, with the same energy density in the initial and final distribution functions. The transport coefficients are $D = 1.2$\,GeV$^2$/fm, $\varv=-2$\,GeV/fm. 
		Equilibrium is reached faster compared to the underoccupied setting of Fig.\,\ref{fig3} and is approached from above in the IR at large times. 
		The dot-dashed curve is the Rayleigh--Jeans, the dotted line the Maxwell--Boltzmann distribution.
	}
\end{figure}
\begin{figure}[t]
	\centering
	\includegraphics[scale=0.80]{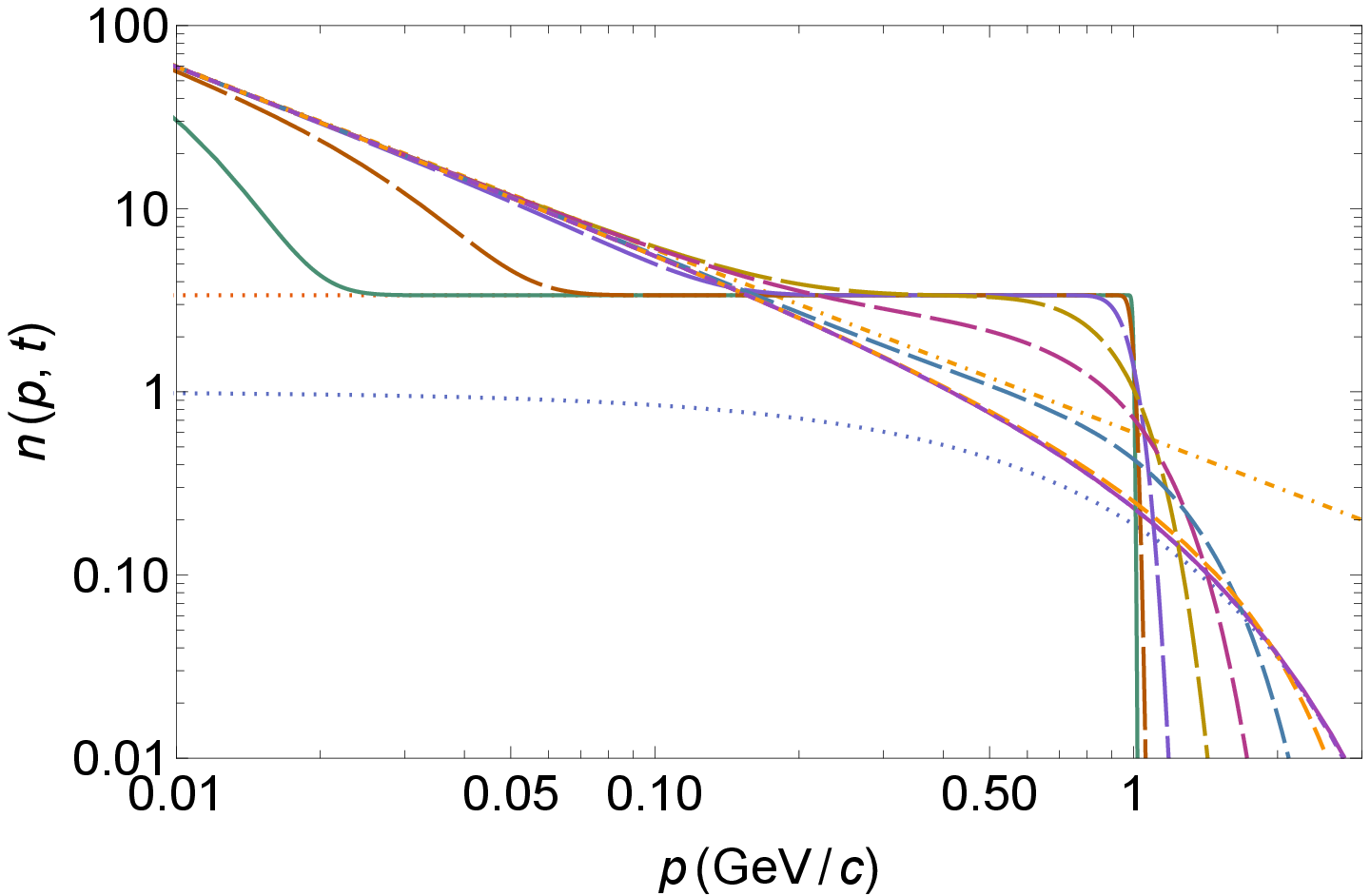}
	\caption{\label{fig6}
		Thermalization of gluons through inelastic collisions in an overoccupied case as in Fig.\, 5, but in a log-log-plot to emphasize the equilibration of the soft modes. The time sequence is as in Fig.\,\ref{fig5} plus two additional time steps, $t=0.4$ and 2 fm/$c$.  The initial occupation corresponds to a final equilibrium temperature $T=600$\,MeV. In the IR, equilibrium is reached faster compared to the underoccupied setting of Fig.\,\ref{fig4}, and approached from above for sufficiently large times.
		The dot-dashed line is the Rayleigh--Jeans, the dotted curve the Maxwell--Boltzmann distribution.
	}
\end{figure}

First the approach to equilibrium through inelastic gluon scatterings with $\mu=0$  in an underpopulated system with an (arbitrary) average initial occupation $n_\text{i}^0=0.12$ is discussed. This is below the critical value $r_\text{c}=0.154$
for a theta-function initial distribution and $Q_\text{s}=1$\,GeV/$c$. Assuming conserved energy density, it corresponds to an equilibrium temperature  $T=-D/\varv\simeq267$\,MeV, which is far below typical initial central temperatures at LHC energies. For an equilibration time at $p=Q_\mathrm{s}$ of 
$\tau_\text{eq}=4D/(9\varv^2)\simeq 0.14$ fm/$c$,
the corresponding transport coefficients as obtained from Eqs.\,(\ref{Dv}) are $D = 0.23$\,GeV$^2$/fm, $\varv=-0.86$\,GeV/fm.

The resulting solutions of the NBDE for the gluon distribution functions are shown in Fig.\,\ref{fig3} at $t =2\times10^{-5}, 2\times10^{-4}, 2\times10^{-3}$, $0.01, 0.04$, and $0.12$\,fm/$c$, with decreasing dash lengths. At $p=0$, thermalization is instantaneous. For $p=Q_\mathrm{s}$, the last curve ($t=0.12$\,fm/$c$) has almost reached equilibrium, whereas thermalization  at low momenta $p\lesssim0.1$\,GeV/$c$ takes much less time, $\tau_{\infty}(p\ll Q_\mathrm{s})\ll \tau_\mathrm{eq}$.
The soft modes always approach equilibrium from below, and quickly within $t\simeq 10^{-3}$\,fm/$c$ reach the Rayleigh--Jeans approximate form $n_\mathrm{RJ}(p)=T/p$ of the thermal distribution (dot-dashed curve).

The steep cutoff in the UV at $p = Q_\mathrm{s}$ is rapidly smeared out at short times --- this was already the case in the free solution without boundary conditions \cite{gw18}. 
The dotted line is the Maxwell--Boltzmann distribution that is attained in thermal equilibrium at large momenta.
In contrast to the soft modes, much more time is needed to equilibrate the tail towards the thermal distribution, where about $\tau_{\infty}(p>Q_\mathrm{s})\gtrsim 2$\,fm/$c$ is required. This differs from the linear relaxation ansatz, where the thermal tail is reached very fast in the full large-momentum range due to the enforced exponential approach to equilibrium.
The NBDE solutions of Fig.\,\ref{fig3} are shown in Fig.\,\ref{fig4} in a log-log-plot to emphasize the fast thermalization of the low-momentum modes through inelastic collisions.

The effect of overpopulation as compared to the underpopulated case in Fig.\,\ref{fig3} is shown in Fig.\,\ref{fig5}, where the initial occupation is enhanced to $n_\text{i}^0=3.37$ to comply with a final equilibrium temperature $T=600$\,MeV that is obtained assuming the same energy density in the initial and final distribution functions. In regions of high initial relative occupancy, $n\,(p,t)$ is reduced in the course of the time evolution towards the thermal value through gluon merging. Equilibrium is again reached instantaneously for $p=0$, and faster in the IR compared to the underoccupied case of Fig.\,\ref{fig3}. However, thermalization in the UV tail remains very similar.
Fig.\,\ref{fig6} shows the same calculation as in Fig.\,\ref{fig5}, but in a log-log-plot that emphasizes the fast equilibration of the soft modes through gluon splitting in the overoccupied case that is realistic at LHC energies. 

These closed-form results from a rather schematic nonlinear diffusion equation are surprisingly similar to previous numerical solutions \cite{jpb13} of transport equations in the soft sector that contain dynamical integrals over distribution functions that themselves are evolving in time. This is particularly evident when comparing with Fig.\,8 in Ref.\,\cite{blmt17} for $n_\mathrm{i}^0=1$ and inelastic collisions: When physical timescales are inserted into their results, these are hardly distinguishable from NBDE-solutions for $n_\text{i}^0=1$ and $\tau_\mathrm{eq}=0.14$\,fm/$c$. (For a NBDE-calculation with $\tau_\mathrm{eq}=0.25$\,fm/$c$ that was proposed as an average relaxation time in Ref.\,\cite{fmr18}, the agreement is still reasonable, but slightly less convincing.) This seems to indicate that it is permissible to simplify the nonperturbative relativistic many-body problem using mesoscopic nonlinear partial differential equations that capture the essence of the time evolution of thermalization, even though they do not account for every detail of the thermalization process that is governed by nonperturbative QCD: The intricate QCD evolution is hidden in the transport coefficients.

Equilibration at $p=0$ via inelastic collisions through gluon merging or splitting is instantaneous --- unless a finite gluon formation time is introduced. In contrast, thermalization at $p=0$ through elastic collisions with $\mu_\mathrm{i}<0$ takes a finite time until 
$\mu=0$ is reached \cite{sgw21}, and gluons move into the condensed state in case of overoccupation. Hence, this argument confirms that condensate formation is essentially forbidden, or at least strongly hindered \cite{blmt17} in non-Abelian systems.
\section{Conclusions}
Thermalization of massless gluons in relativistic hadronic collisions has been investigated based on the nonlinear boson diffusion equation, which is an approximate kinetic equation that has the Bose--Einstein distribution as a stationary limit. It has been derived earlier from a quantum Boltzmann collision term with an energy-conserving function that has a finite width. The time-dependent solutions can be calculated in the full momentum range provided the boundary conditions at the singularity $p=\mu$ are properly considered. Drift and diffusion functions $\varv\,(p,t)$, $D\,(p,t)$ contain the complexity of the partonic many-body system as accounted for by nonperturbative QCD. 

With the simplifying assumption of constant transport coefficients, the NBDE has been solved in closed form through the logarithmic derivative of a time-dependent partition function that comprises an integral over the initial conditions and Green's function. As a significant improvement of the analytical model, boundary conditions at the singularity have been included, thus causing the correct Bose--Einstein limit not only in the thermal tail, but also for the low-momentum modes. Schematic initial conditions as proposed in the colour-glass model for the nonequilibrium gluon distribution at $t=0$ have been used, but it would be easy to test other initial distributions within the model.

The quantum-statistical properties of bosons with a singularity at $p=\mu\le0$ are essential for the boson enhancement and the thermalization at low momenta.  The non-Abelian properties of gluons mediate inelastic collisions through splitting and merging such that the particle number is not conserved, the chemical potential vanishes, and the system thermalizes very quickly at $p=0$ for both over- and underoccupied conditions. In contrast, thermalization at $p=0$ through number-conserving elastic collisions would require a finite time to increase the chemical potential towards 
$\mu=0$, where particles could start to move into the condensed state.

Hence, gluon condensation is not possible in the NBDE-model because thermalization at $p=0$ through inelastic collisions occurs instantaneously due to the boundary condition --- which may, however be modified if the quantum-mechanical time for gluon splittings or mergings would be taken into account.
Indeed, in the numerical model of Ref.\,\cite{blmt17} a small time delay is observed also for inelastic collisions, a competition occurs between elastic and inelastic processes not only at finite momenta, but also at $p=0$, and transient condensate formation could occur.

At the gluon saturation scale, the local thermalization time $\tau_{\infty}\,(p=Q_\mathrm{s})$ in the time-dependent solutions is found to agree with the analytical expression 
$\tau_\mathrm{eq}=4D/(9\varv^2)$ that had been derived for an initial occupation $n_\text{i}^0=1$ from an asymptotic expansion at $p=Q_\mathrm{s}$ of the error functions occurring in the exact solutions of the free case \cite{gw18}. Thermalization is reached much faster in the soft modes and more slowly in the thermal tail, in agreement with investigations that are more directly based on QCD-considerations.

The results for inelastic scattering are rather similar to what had been obtained before by other groups \cite{jpb13,blmt17} in numerical solutions of transport equations that contain dynamical integrals over distribution functions that themselves are evolving in time. This suggests that the schematic NBDE-approach to the thermalization process captures already essential ingredients of the problem.
Hence, 
the nonlinear boson diffusion model offers basic insights into the physics of thermalization
that can otherwise only be obtained through substantial numerical efforts. As in related kinetic thermalization theories, gluons are taken to be massless with $\epsilon=\sqrt{(p^2+m^2)}\rightarrow p$. In case of a finite gluon mass as suggested in quasiparticle models, modifications of the occupation-number distributions may be expected in the IR, but the UV tails are essentially unaffected. Since in the ergodic approximation, the singularity appears in energy space, it would still have to be treated explicitly for finite gluon mass.

It remains to be checked whether the instantaneous equilibration of the $p=0$ mode through inelastic collisions already describes the physics correctly, or has to be modified through the consideration of a gluon splitting or fusion time.
 The dependence on different initial occupations as well as more realistic initial conditions should also be investigated. For an ambitious extension of the model, the assumption of constant transport coefficients could be relaxed, but it is unlikely that closed-form solutions can then still be obtained. A derivation of drift and diffusion functions from QCD would be most welcome.
\bigbreak
\noindent
\noindent
\textbf{Acknowledgements}\\
I thank Johannes H\"olck (ITP Heidelberg) for discussions, 
colleagues and referees for constructive comments, including sceptical ones from a QCD-viewpoint.
\bibliographystyle{elsarticle-num}

\end{document}